\documentclass{aa}
\usepackage{graphics,psfig}
 
\begin{document}      

\title{The frequency evolution of interstellar pulse broadening
from radio pulsars}

\author{
        O.~L\"ohmer \inst{1},
        D.~Mitra \inst{1},
        Y.~Gupta \inst{2},
        M.~Kramer \inst{3},
        A.~Ahuja \inst{4}
        }

\institute{Max-Planck-Institut f\"ur Radioastronomie, Auf dem H\"ugel 69,
              D-53121 Bonn, Germany
    \and
           National Centre for Radio Astrophysics, TIFR,
           Pune University Campus, Ganeshkhind,
           Pune 411007, India  
    \and
           University of Manchester, Jodrell Bank Observatory,
           Macclesfield, Cheshire SK11 9DL, UK
    \and
           Inter-University Centre for Astronomy and Astrophysics,
           Pune University Campus, Ganeshkhind, Pune 411007, India
           }

\offprints{O.~L\"ohmer, e-mail: loehmer@mpifr-bonn.mpg.de}
\date{Received / Accepted}

\titlerunning{Frequency evolution of interstellar pulse broadening}
\authorrunning{O.~L\"ohmer et al.}

\abstract{ In this paper we report multi-frequency measurements of
  pulse broadening times ($\tau_{\rm d}$) for nine medium dispersion
  measure (DM $\approx 150-400$ pc cm$^{-3}$) pulsars observed over a
  wide frequency range. The low frequency data at 243, 325 and 610 MHz
  are new observations done with the Giant Metrewave Radio Telescope
  (GMRT). The frequency dependence of $\tau_{\rm d}$ for all but one
  (\object{PSR B1933+16}) of our sources is consistent with the
  Kolmogorov spectrum of electron density fluctuations in a turbulent
  medium. PSR~B1933+16, however, shows a very flat spectrum as
  previously observed for high DM pulsars. Our observations combined
  with earlier published results enable us to study the spectral index
  of $\tau_{\rm d}$ over the whole observed DM range. While the
  spectral properties are generally consistent with a Kolmogorov
  spectrum, pulsars seen along line-of-sights towards the inner Galaxy
  or complex regions often show deviations from this expected
  behaviour.  }

\maketitle

\section{Introduction
\label{intro}}

The free electron density distribution in the interstellar medium
(ISM) can be decomposed into three different regimes: the spiral arms
consisting of gaint HII regions, the inner disk comprising
of a dense ionized region and the thick disk filled with diffuse
electron gas (Taylor \& Cordes 1993\nocite{tc93}). The pulsar signal
traversing the ISM suffers interstellar dispersion quantified in terms
of dispersion measure, DM, which is the integrated electron column
density towards the pulsar (at distance $D$), i.e.\
$\rm{{DM}}=\int_{0}^{D} n_e\;dl$.  Fluctuations of the electron
density along the line of sight (LOS) give rise to several observable
scattering effects, which manifest themselves as observational
properties like angular broadening, temporal pulse broadening and
scintillation of pulsars (see Rickett 1990\nocite{ric90} for a
review).

Scattering causes propagation of signals along a variety of different
ray paths with different geometrical lengths, so that a pulse, which
has left the pulsar at one instant, arrives at the observer over a
finite time interval.  For a gaussian distribution of irregularities
and applying the thin screen approximation, the pulse broadening
function (PBF) of the ISM can be well described by an exponential
decay of the pulse, i.e. PBF$(t) = {\rm exp}(-t/ \tau_{\rm d})$, where
$\tau_{\rm d}$ is called the pulse broadening time. (Scheuer 1968).
Random interference among the different paths produces a diffraction
pattern in the plane of the observer. This pattern decorrelates over a
characteristic bandwidth $\Delta\nu_{\rm d}$. Both $\tau_{\rm d}$ and
$\Delta\nu_{\rm d}$ strongly depend on frequency $\nu$ (i.e.
$\tau_{\rm d} \propto \nu^{-\alpha}$ and $\Delta\nu_{\rm d} \propto
\nu^{ \alpha}$) and are related to each other as
\begin{equation}\label{uncertain}
2\pi\,\tau_{\rm d}\,\Delta\nu_{\rm d} = C_1\ \ .
\end{equation}
The constant $C_1$ is of the order unity but changes for different
models and geometries for the electron density fluctuations (e.g.
Lambert \& Rickett 1999\nocite{lr99}). However, until now there exists
no direct measurement of $\tau_{\rm d}$ and $\Delta\nu_{\rm d}$ at the
same observing frequency for obtaining $C_1$.

The strength of scattering of radio waves caused by electron density
fluctuations in the ISM has been subject of detailed studies since the
discovery of scintillation in pulsar signals (Scheuer
1968\nocite{sch68}). A commonly used description of the scattering
strength is to attribute a power law to the electron density spectrum
with a large range between ``inner'' and ``outer'' scales,
$k_{i}^{-1}$ and $k_{o}^{-1}$; i.e.,
\begin{equation}
P_{n_e} (q) = \frac{C_{n_e}^{2}}{(q^{2} + k_{o}^{2})^{\,\beta/2}}\
              {\rm{exp}} \left(-\frac{q^{2}}{4\,k_i^2}\right)\ \ ,
\end{equation}
where $q$ is the magnitude of the three-dimensional wave\-number and
$C_{n_e}^2$ is the fluctuation strength for a given LOS (e.g. Rickett
1977\nocite{ric77}). For $k_{o} \ll q \ll k_{i}$, one obtains a simple
power-law model with a spectral index $\beta$, i.e., $P_{n_e} (q) =
C_{n_e}^{2} q^{-\beta}$ and $\alpha = 2 \beta /(\beta -2)$. For a pure
Kolmogorov spectrum of density irregularities, $\beta = 11/3$, we
expect $\alpha = 4.4$. Attempts to reconstruct the electron density
spectrum from various observations (Armstrong et al.\
1995\nocite{ars95}) have led to the suggestion that over 5 decades of
wavenumber range 10$^{-13}$ m$^{-1} <$ q $< 10^{-8}$ m$^{-1}$ the
spectrum can be approximated by a power law with an average index
$\langle\,\beta\,\rangle\sim 3.7$, close to the value expected from
the Kolmogorov process. However, there is some evidence for deviations
from this picture, over some parts of the wavenumber range (see, for
example, Gupta 2000\nocite{gup00b} and reference therein).

Furthermore, it is uncertain how the scattering material and processes
operate in different directions of the Galaxy. Multi-frequency scatter
broadening measurements have been used to determine $\alpha$ for
several LOSs in the Galaxy. For LOSs with low DM's ($< 50$~pc
cm$^{-3}$) there is typically good agreement with the Kolmogorov
spectrum (Cordes et al. 1985\nocite{cwb85}; Johnston et al.\
1998\nocite{jnk98}). This is also supported by estimates of $\alpha$
in the local ISM using diffractive and refractive interstellar
scintillation (ISS) observations of pulsars (Bhat et al.\
1999\nocite{bgr99}). These local LOSs are most probably sampling the
diffuse, homogeneous electron density component of the Galaxy, where
the Kolmogorov process hold true. High DM ($> 400$~pc cm$^{-3}$)
pulsars towards the inner parts of the Galaxy, in contrast, have an
average $\langle\,\alpha\,\rangle = 3.44 \pm 0.13$, showing
significant departures from the Kolmogorov spectrum (L\"ohmer et al.\
2001\nocite{lkm+01}, hereafter Paper I). We interpreted this
phenomenon in Paper I as anomalous scattering as discussed by Cordes
\& Lazio (2001\nocite{cl01}).  In this picture the observed anomalous
behaviour can be explained by invoking scattering caused by multiple
scattering screens with anisotropic irregularities and finite
transverse extent along the LOS. As a consequence, less radiation
reaches the observer at lower frequencies since some of the radiation
that would be scattered by an infinite screen is now lost, causing
$\alpha$ to be lower than the standard Kolmogorov value. Indeed,
towards these directions at low Galactic latitudes one encounters
numerous HII regions embedded in the dense thin disk, so that the
probability to find such multiple screens along the LOS is quite high.

In order to get a better understanding of the scattering properties in
the ISM we need measurements of $\alpha$ for more LOSs. In particular
the scattering properties of intermediate DM pulsars ($100\,\la\,
{\rm DM}\,\la\, 400$~pc cm$^{-3}$) have not been studied yet. In an
effort to find $\alpha$ for intermediate DM pulsars we have currently
launched an observational program using the Gaint Metrewave Radio
Telescope (GMRT) in Pune, India. In this paper we report results for
the first phase of these observations.

\section{Observations and Data analysis \label{obs}}

The following selection criteria were used to find an adequate sample
of intermediate DM pulsars for our GMRT observations. Firstly pulsars
with flux density $>$ 10 mJy at 400 MHz were chosen from the Taylor et
al.\ (1993; updated version 1995\nocite{tml93})
catalogue.  We also ensured that the expected width of the pulse,
i.e. the intrinsic width along with the pulse broadened width, at
the lowest frequency (243 MHz) is smaller than 80\% of the pulse
period. Using an intrinsic pulse width of 5\% of the period at
$\sim$5~GHz we estimated the width at lower frequencies according to a
power-law index of $\lambda^{0.25}$ (e.g.\ Thorsett
1991\nocite{tho91a}), where $\lambda$ is the wavelength in meters. We
found the expected pulse broadening at each frequency by applying
the empirical relation $\tau_{\rm d}\ {\rm (ms)} = 4.2\times
10^{-5}~DM^{1.6} \times\;(1+3.1\times 10^{-5}~DM^{3})\;\lambda^{4.4}$
(Ramachandran et al.\ 1997\nocite{rmdm+97}). The final estimated
width was found by adding the two widths in quadrature. Applying the
above selection criteria we ended up with a sample of 33 pulsars.
Here we report results of the first phase of GMRT observations of nine
pulsars.

The observations were carried out with the GMRT in February 2002 using
three bands around center frequencies at 243, 325 and 610~MHz. The
GMRT has a `Y' shaped hybrid configuration of antennas with 14
antennas placed randomly in a compact central array of 1 km by 1 km,
and the remaining 16 antennas distributed along the three arms of the
`Y' (Swarup et al.\ 1997\nocite{sas+97}). The GMRT was in its
commissioning phase during our observations, and due to various
maintenance activities not all of the 30 antennas were available for
observations. The observations were carried out with typically 20 to
25 antennas.  Dual circular polarization signals from all the selected
antennas were incoherently added (i.e., signals from each antenna were
first detected and then added) in a 256 channel filter bank with a
total bandwidth of 16 MHz.  The summed signals were integrated to a
time constant of 0.516 ms and were recorded for off-line analysis
after adding the two polarizations (see Gupta et al.\
2000\nocite{ggj+00} for more details about the pulsar mode of
operation of the GMRT). The filterbank characteristics result in
dispersion smearing per channel\footnote{$t_{\rm DM} = 8.3~{\rm
DM}~\Delta \nu_{\rm BW} / \nu^3\: \mu s$, where DM is the dispersion
measure (pc cm$^{-3}$), $\nu$ is the observing frequency (GHz) and
$\Delta \nu_{\rm BW}$ the filterbank channel bandwidth (MHz).},
$t_{\rm DM}$, for the pulsars observed of 5.7~ms $ \le t_{\rm DM} \le
$~14.5~ms at 243~MHz, of 2.4~ms $ \le t_{\rm DM} \le $~6.1~ms at
325~MHz, and of 0.4~ms $ \le t_{\rm DM} \le $~0.9~ms at 610~MHz.

Typical observation times were between 10 and 45~min, depending on the
observing frequency and flux density of the individual source.  During
off-line reduction the filterbank signals were de-dispersed and
summed.  We analysed the signals for RFI and skipped those parts of
the data with spikes larger than a factor of three of the rms of the
data. The signals were folded with the topocentric pulse period to
produce total power profiles.

\begin{figure}
\centering
\psfig{file=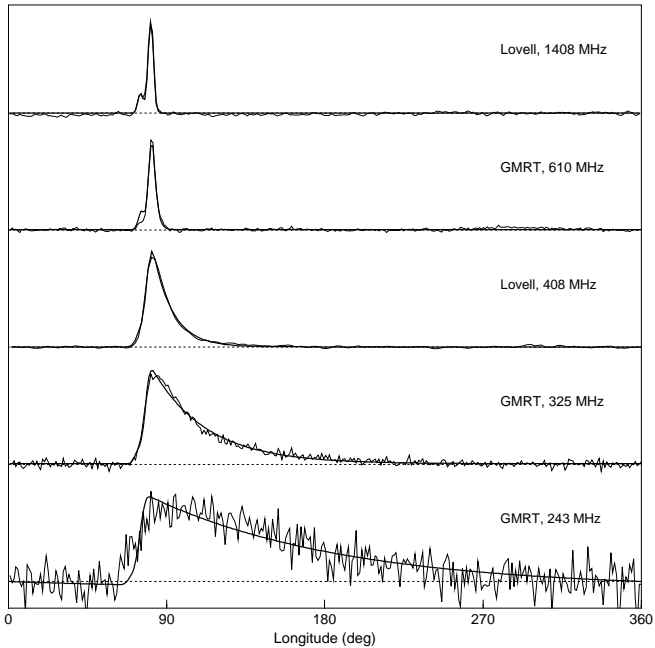,width=8cm}
\caption[]{ \label{fig:1831} Integrated pulse profiles and best-fit
model profiles for PSR B1831$-$03 at different frequencies. The
profiles at 243, 325 and 610~MHz were observed with the GMRT, whereas the
408 and 1408 MHz~profiles were taken from the EPN database (Lovell
observations). The alignment of the profiles for different frequencies
was done with respect to the peak of the main pulse.  }
\end{figure}
%
At 408 and 1408 MHz we used pulse profiles\footnote{These profiles are
publicly available from the European Pulsar Network (EPN) archive
maintained by the MPIfR, Bonn,
http://www.mpifr-bonn.mpg.de/div/pulsar/data/} observed with the 76~m
Lovell telescope at Jodrell Bank, UK. Using both circular
polarizations the signals were incoherently de-dispersed and added in
filterbanks with varying bandwidths. At 408~MHz 32 channels of
0.125~MHz bandwidth (2.4~ms $ \le t_{\rm DM} \le $~6.0~ms) were used,
and at 1408~MHz 32 channels of 1~MHz each (0.5~ms $ \le t_{\rm DM} \le
$~1.2~ms) were employed.  Details of the system can be found in Gould
\& Lyne (1998)\nocite{gl98}.

In order to measure the pulse broadening time $\tau_{\rm d}$ we used
the method described in Paper I. Firstly, we construct a pulse
template as a sum of Gaussian components fitted to an observed high
frequency profile, where the scattering for the given pulsars is
negligible. For our sample, the 1408 MHz profiles were adequate. For
each lower frequency, we then find $\tau_{\rm d}$ from the best fit of
the model profile, which is the convolution of the template with the
dispersion smearing and the adopted PBFs, to the observed profile.
The exact functional form for the PBF of the ISM is not known. We thus
analyse the fits for three trial PBFs, (1) the PBF for a thin screen
(PBF$_1$) and (2) for a uniformly distributed medium (PBF$_2$) in an
ISM with Gaussian density fluctuations, given by (Williamson 1972,
1973\nocite{wil72}\nocite{wil73}):
\begin{eqnarray}
{\rm PBF}_1(t) &=& {\rm exp}(-t/ \tau_{\rm d})\,U(t)\label{thinscreen}\\
{\rm PBF}_2(t) &=& (\pi^5\tau_{\rm d}^3/8t^5)^{1/2}\: {\rm exp}(-\pi^2\tau_{\rm
d}/4t) \, U(t)\label{unimedium}\ ,
\end{eqnarray}
where $U(t)$ is the unit step function, i.e.\ $U(t<0)=0,
U(t\ge0)=1$. (3) The third PBF is characterized by density
fluctuations with a L{\'e}vy probability distribution function that
has a power-law decay (Boldyrev \& Gwinn 2003\nocite{bg03}) and an
asymptotic form PBF$_3(t)=(t/\tau_{\rm d})^{-4/3}\,U(t)$ for $t/
\tau_{\rm d}\gg 1$.  As for the high DM pulsars presented in Paper I, we
again find PBF$_1$ to be most appropriate to describe the observed
scattering. In particular, the fits using PBF$_2$ and PBF$_3$ cannot
reproduce the long ``scattering tails'' observed at lower frequencies,
resulting in $\chi^2$ values that are larger by factors of 2 and
more. Using the thin screen approximation we obtain best-fit values
and uncertainties for $\tau_{\rm d}$ from the $\chi^2$ contours in the
plane of $\tau_{\rm d}$ and offset in phase.

In Fig.~\ref{fig:1831} observed and best-fit model profiles for the
observed frequencies are shown for \object{PSR B1831$-$03}. The
template is constructed from the 1408~MHz Lovell profile using two
Gaussians. Note the high S/N ratio and quality of the GMRT profiles
proving that this new telescope is highly capable of pulsar
observations at low radio frequencies.  The best-fit model profiles
describe the shape of the observed profiles in an excellent manner. At
610~MHz the small peak at the leading part of the profile was not
observed, which, however, does not affect the $\tau_{\rm d}$
measurement (see next paragraph). At 243~MHz the dispersion smearing
at the leading part of the profile seems to be not adequately
described by the model profile, resulting in a much steeper rise of
the peak.  We repeated the fit using artificially increased dispersion
smearing functions and found that the effect on $\tau_{\rm d}$ is well
below its 1$\sigma$ uncertainty and therefore negligible.

As noted in Paper I, intrinsic profile variations with frequency (see
the 610~MHz profile of Fig.~\ref{fig:1831}) could in principle give
rise to inaccurate estimation of pulse broadening times. A careful
analysis of these effects on the measured $\tau_{\rm d}$ can be done
using simulated pulse profiles with frequency evolution that are made
subject to pulse broadening. As shown, resulting deviations of the
measured $\tau_{\rm d}$ values from the true ones are in fact very
small and can be accounted for using increased error bars. Thus,
we again quote conservative 3$\sigma$ error bars for all scatter
broadening times.

\tabcolsep5pt
\renewcommand{\arraystretch}{1.2}
\begin{table*}
\caption
    {Pulse Broadening Measurements \label{tab:tau}
}
\begin{tabular}{crrccrrrrrrr}
\hline\hline\noalign{\smallskip}
PSR & \multicolumn{1}{c}{$l$} & \multicolumn{1}{c}{$b$} &
\multicolumn{1}{c}{DM} & \multicolumn{1}{c}{D} &
\multicolumn{1}{c}{$\tau_{\rm 243}$} &
\multicolumn{1}{c}{$\tau_{\rm 325}$} & 
\multicolumn{1}{c}{$\tau_{\rm 408}$} &
\multicolumn{1}{c}{$\tau_{\rm 610}$} &
\multicolumn{1}{c}{$\alpha$} & \multicolumn{1}{c}{$\beta$} & 
\multicolumn{1}{c}{$\log C^2_{\rm n_e}$} \\
 & \multicolumn{1}{c}{(deg)} & \multicolumn{1}{c}{(deg)} &
\multicolumn{1}{c}{(pc~cm$^{-3}$)} & \multicolumn{1}{c}{(kpc)} & 
\multicolumn{1}{c}{(ms)} & \multicolumn{1}{c}{(ms)} & 
\multicolumn{1}{c}{(ms)} & \multicolumn{1}{c}{(ms)} & & & \\
\multicolumn{1}{c}{(1)} & \multicolumn{1}{c}{(2)} & 
\multicolumn{1}{c}{(3)} & \multicolumn{1}{c}{(4)} & 
\multicolumn{1}{c}{(5)} & \multicolumn{1}{c}{(6)} & 
\multicolumn{1}{c}{(7)} & \multicolumn{1}{c}{(8)} & 
\multicolumn{1}{c}{(9)} & \multicolumn{1}{c}{(10)} & 
\multicolumn{1}{c}{(11)} & \multicolumn{1}{c}{(12)}\\
\noalign{\smallskip}\hline\noalign{\smallskip}
\object{B1821$-$19} & 12.3 & $-3.1$ & 224 & 4.7 &\multicolumn{1}{c}{...} &
58 (23) & 22 (4) & 4.6 (0.2) & $4.0_{-0.7}^{+0.5}$ & 
$4.0_{-0.4}^{+1.1}$ & -1.1\\
\object{B1826$-$17} & 14.6 & $-3.4$ & 218 & 4.7 &\multicolumn{1}{c}{...} &
137 (37) & 41 (6) & 7.5 (1.3) & $4.6_{-0.5}^{+0.5}$ &
$3.6_{-0.2}^{+0.4}$ & -0.9\\
B1831$-$03 & 27.7 & $2.3$  & 236 & 5.2 & 208 (94) & 59 (4) &
21.3 (1.1) & 3.2 (0.5) & $4.5_{-0.6}^{+0.4}$ &
$3.6_{-0.2}^{+0.5}$ & -1.2\\
\object{B1845$-$01} & 31.3 & $0.0$  & 159 & 4.0 &\multicolumn{1}{c}{...} &
262 (181) & 54 (8) & 17.0 (1.4) & $4.3_{-1.2}^{+0.7}$ &
$3.8_{-0.4}^{+1.9}$ & -0.5\\
\object{B1859+03}   & 37.2 & $-0.6$ & 401 & 7.3 &\multicolumn{1}{c}{...} &
241 (51) & 89 (9) & 12.6 (0.9) &  $4.7_{-0.4}^{+0.3}$ &
$3.5_{-0.1}^{+0.2}$ & -1.0\\
\object{B1900+01}   & 35.7 & $-2.0$ & 246 & 3.4 & 176 (58) & 34 (2) &
14.7 (1.3) & $< 1.0$ & $4.8_{-0.8}^{+0.6}$ &
$3.4_{-0.2}^{+0.6}$ & -1.0\\
\object{B1920+21}   & 55.3 & $2.9$  & 217 & 7.6 & 4.5 (1.5) & $< 0.8$ &
$< 1.7$ & $<0.6$ & \multicolumn{1}{c}{...} &
\multicolumn{1}{c}{...} & -2.9\\
B1933+16   & 52.4 & $-2.1$ & 159 & 5.6 & 4.6 (0.2) & 1.8 (0.1) & 
$< 1.9$ & $< 0.4$ & $3.4_{-0.2}^{+0.2}$ &
\multicolumn{1}{c}{...} & -2.7\\
\object{B2002+31}   & 69.0 & $0.0$  & 235 & 7.5 & 27 (13) & 5.3 (1.2) &
2.0 (1.4) & $< 0.7$ & $5.0_{-1.5}^{+1.7}$ &
$3.3_{-0.5}^{+1.3}$ & -2.3\\
\noalign{\smallskip}\hline
\end{tabular}
\vspace{0.2cm}\\ {\small Note.-- Cols.~(2) and (3) give the Galactic
longitude and latitude of each pulsar; col.~(4) its Dispersion Measure
(DM) and col.~(5) its DM distance (from the Cordes \& Lazio (2002)
model, hereafter NE2001); cols.~(6)--(9) the pulse broadening times
with 3$\sigma$ errors in parenthesis at 243, 325, 408 and
610~MHz. Upper limits are quoted with ``$<$''. Col.~(10) gives the
spectral index of pulse broadening and col.~(11) the spectral index
of density irregularities (using $\beta = 2\alpha/(\alpha-2)$) for
each pulsar with their 1$\sigma$ errors. Col.~(12) gives the
logarithm of the fluctuation strength $C^2_{\rm n_e}$.  }
\end{table*}
%

Recently, another method to analyse pulse broadening related to the
CLEAN alogrithm was proposed by Bhat et al.\ (2003\nocite{bcc03}). In
their approach, the authors try to derive the intrinsic pulse shape at
the observed frequency without using any knowledge of the pulse
profile at another, higher frequency. They point out that utilizing a
high frequency template can indeed lead to uncertainties due to the
same unkown frequency evolution of the pulse profile that we try to
simulate in our computations (see Paper I).  Whilst it is indeed more
straightforward in their method to perform a deconvolution to recover
the intrinsic profile, their alogrithm cannot always produce unique
results, yielding strikingly different values and hence uncertainties,
sometimes.  This is demonstrated for \object{PSR B1849+00} which was
also studied in Paper I. Applying PBF$_1$ and PBF$_2$ (see
Eq.~\ref{thinscreen} and \ref{unimedium}) the authors obtain equally
good fits for $\tau_{\rm d}=225\pm14$~ms and $\tau_{\rm
d}=121\pm6$~ms, where a choice can only be made by making an
assumption about the more likely intrinsic profile.  A comparison of
these values with our measurement of $\tau_{\rm d}=223\pm24$~ms as
derived in Paper I shows that both methods result in consistent pulse
broadening times for the case of the thin screen approximation. This
supports our findings that an exponential decay is the most
appropriate form to describe pulse broadening for intermediate and
high DM pulsars. The example of PSR B1849+00 shows that extra, a
priori information (typically an idea of the expected pulse shape) is
usually needed to obtain correct solutions for more complicated
profiles which holds true for both the CLEAN algorithm as well as our
approach.  Given the apparent imperfections of both methods, all
derived values should be treated with considerable care, e.g.~by
reflecting the possible systematic errors by increasing the error
estimates correspondingly, as done in our study. It is comforting to
note that for PSR B1849+00 the frequency dependence of $\tau_{\rm d}$,
derived by Bhat et al.\ (2003, $\alpha=3.5\pm0.7$), and us (Paper I,
$\alpha=2.8^{+1.0}_{-0.6}$) are consistent. Recent OH observations
toward PSR B1849+00 revealed absorption features that most likely
originate from a small and dense molecular clump (Stanimirovi{\' c} et
al.\ 2003\nocite{swd+03}). Thus, the LOS to the pulsar probes complex
material so that our findings of non-Kolmogorov frequency dependence
of pulse broadening is not surprising.

\section{Results and Discussion \label{results}}

Table~\ref{tab:tau} summarises the multi-frequency measurements of
pulse broadening times for our sample of pulsars. PSRs B1821-19
and B1859+03 were not observed at 243~MHz due to the limited telescope
time. For PSRs B1826$-$17 and B1845$-$01 we did not detect a pulse
profile at this frequency, probably because the pulse is smeared due
to high scattering.

\begin{figure}[t]
\centering
\psfig{file=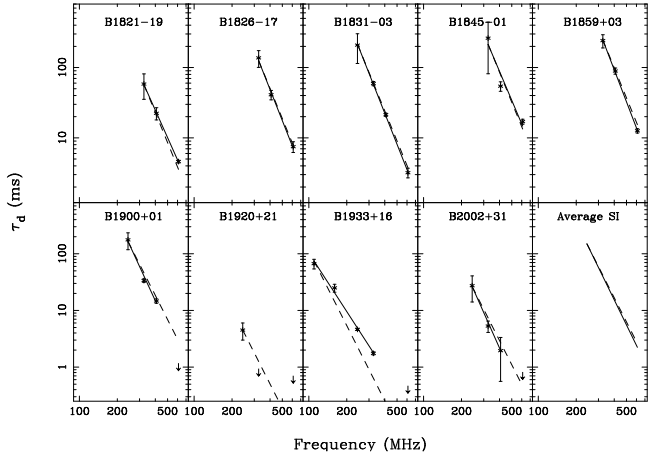,angle=-90.,width=8.5cm}
\caption{ \label{fig:scatfreq} Pulse broadening times, $\tau_{\rm d}$,
with their 3$\sigma$ errors as a function of observing frequency, $\nu$,
for nine pulsars. Arrows denote measured upper limits on $\tau_{\rm
d}$. The lines correspond to the linear fit of the form $y = -\alpha\,
x + K$ where $y = \log(\tau_{\rm d})$ in ms and $x = \log(\nu)$ in
MHz. The dashed lines are examples of the expected dependence due to a
Kolmogorov spectrum, i.e. $\alpha = 4.4$.  The bottom right most panel
shows a linear function with the derived average spectral index
$\langle \alpha \rangle = 4.57$. See text for further details.  }
\end{figure}

\begin{figure}[t]
\centering
\psfig{file=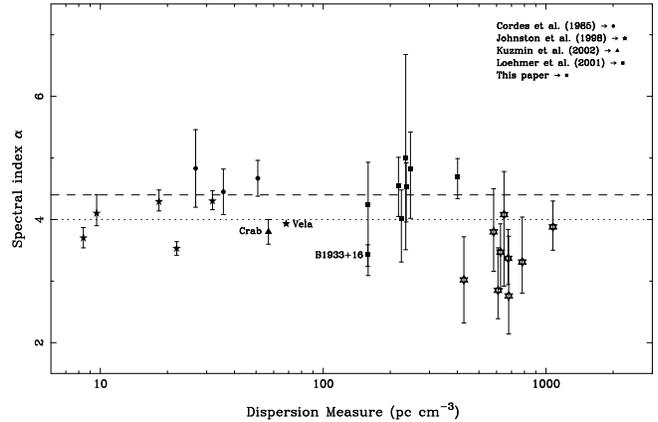,angle=-90,width=8.5cm}
\caption[]{ \label{fig:sidm} Spectral index of pulse broadening,
$\alpha$, as a function of dispersion measure, DM, for the current
sample as well as for earlier measurements\nocite{kkk+02}. The dashed
line $\alpha = 4.4$ indicates the spectral index for a Kolmogorov
spectrum. The dotted line $\alpha = 4.0$ represents the spectral index
for a Gaussian distribution of irregularities.  }
\end{figure}

Fig.~\ref{fig:scatfreq} shows the measured frequency dependence of
$\tau_{\rm d}$. For all pulsars with more than one $\tau_{\rm d}$
measurement, i.e.\ for all but PSR B1920+21, we calculated the
spectral index of pulse broadening, $\alpha$, from Monte-Carlo
simulations as described in Paper I.  For PSR B1933+16, we used values
of $\tau_{\rm d}=67(13)$~ms at 110~MHz and $\tau_{\rm d}=25(4)$~ms at
160~MHz from the literature (Rickett 1977\nocite{ric77}; Slee et al.\
1980\nocite{sdo80}).  The median $\alpha$ and its 1$\sigma$
errors are listed in Table~\ref{tab:tau}. Also listed are the spectral
indices of the electron density spectrum, $\beta$, which we calculated
from the relation $\beta = 2\alpha/(\alpha-2)$. Note that this
relation is only valid for $\beta<4$, i.e.\ for all our pulsars except
PSR B1933+16. The average spectral index for seven pulsars (excluding
PSRs B1920+21 and B1933+16) is $\langle\,\alpha\,\rangle = 4.57 \pm
0.09$.  Fig.~\ref{fig:sidm} shows the spectral index of pulse
broadening as a function of DM for our measurements as well as
published data.  All of our new spectral indices, except the one for
PSR B1933+16, are consistent within their errors with the Kolmogorov
value of 4.4, as found for most of the low DM pulsars as
well. However, along several LOSs, as towards the Crab and Vela
pulsar, PSR B1933+16 and the high DM pulsars, the Kolmogorov theory
fails leading to flattening of the spectra of scatter broadening. The
results for the Crab and Vela pulsar are not surprising, as the
complex structure of the surrounding supernova remnants can introduce
a number of effects affecting the observed scattering properties
(e.g.\ Backer et al.\ 2000\nocite{bwv00}; Lyne et al.\
2001\nocite{lpg01}).

%

The spectral index of pulse broadening for the LOS to PSR B1933+16 is
$\alpha = 3.4(2)$, i.e.\ significantly lower than the Kolmogorov value
of 4.4.  This measured flattening of the spectrum has been observed
before for high DM pulsars and can be explained by anomalous
scattering at multiple scattering screens with finite extensions
and/or varying scattering strength (see Paper I).  Our pulse
broadening measurements for PSR B1933+16 along with published
$\tau_{\rm d}$'s at 110 and 160~MHz fit excellently to a power-law
model with $\log(\tau_{\rm d} {\rm (msec)}) = -3.4\, \log(\nu) + 8.9$,
where $\nu$ is in MHz (as seen in Fig.~\ref{fig:scatfreq}).  It should
be noted that combining measurements at different epochs can be
affected by refractive scintillations which may alter $\tau_{\rm d}$
between epochs and thereby alter values of $\alpha$. However, our
$\tau_{\rm d}$ measurements being quasi simultaneous (separated by
days) are unlikely to be affected by refractive scintillations and the
spectrum derived by these values is in good agreement with the low
frequency values obtained by Cordes et al.\ (1985) at an earlier
epoch. The authors quote $\Delta \nu_{\rm d}$ values for this pulsar
near closeby frequencies of 1.41, 1.42, and 1.67 GHz as 0.125, 0.100,
and 0.110 MHz, respectively. We extrapolated our $\tau_{\rm d}$
spectrum to these frequencies, solved Eq.~\ref{uncertain} for $C_1$
and found it to be 9.1, 7.1, and 4.5, respectively, i.e. much larger
than unity.  The $\Delta \nu_{\rm d}$ measurements could have been
biased by refractive ISS (e.g.\ Gupta et al.\ 1994\nocite{grl94})
leading to an underestimation of the true bandwidths. The resulting
smaller values of $\tau_{\rm d}$ at these frequencies, however, would
lead to a steeper spectrum or to higher $C_1$ values making the
discrepancy even worse\footnote{ Note that there is a possibility that
the nature of the PBF may evolve from the well established exponential
form, in the regime where $\Delta \nu_{\rm d}$ is the easier quantity
to measure.  However, there is no evidence of this from the main
theoretical models (e.g. Lambert \& Rickett 1999 and references
therein).}. Our measured $C_1$'s are much greater than unity and
therefore in contradiction to what is predicted by standard theories
of the ISM.  Lambert \& Rickett (1999), for instance, tabulated
possible values for different geometries and spectral models and found
values ranging from 0.56 to 1.53, i.e.~still much smaller than those
for PSR B1933+16. However, if we were to apply a Kolmogorov spectrum
($\alpha=-4.4$) -- even though it is a very poor fit to the data --
the extrapolated $\tau_{\rm d}$'s for PSR B1933+16 would be much
smaller at the corresponding frequencies, resulting in $C_1$'s that
are much smaller, eventually even smaller than unity.  Thus, the LOS
to PSR B1933+16 remains to be an unresolved puzzle. Careful
observations of decorrelation bandwidths of this pulsar in the
frequency range of 300$-$1400~MHz are needed to help understanding the
properties of the ISM along this LOS.

%
%

\begin{figure}
\centering
\psfig{file=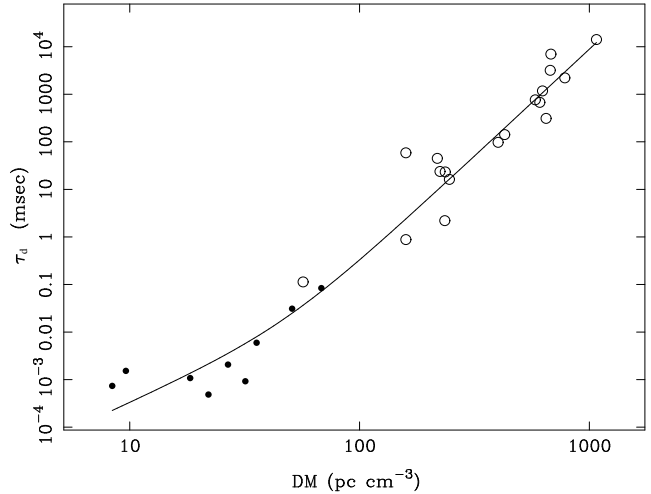,angle=-90,width=8.5cm}
\caption[]{ \label{fig:taudm} Pulse broadening times, $\tau_{\rm d}$,
versus dispersion measure, DM, for 27 pulsars as plotted in
Figure~\ref{fig:sidm}. Filled circles corresponds to decorrelation
bandwidth measurements and open circles are pulse broadening
measurements. The solid line corresponds to the best fit given by
Eq.~\ref{eqn:taudm}. See text for further details.}
\end{figure}

One of the conventional methods of examining the scattering properties
along various LOSs is to investigate the relation between the measured
pulse broadening and the DM. Considering DM as proportional to the
distance $D$ to the pulsar, the power law model of density
irregularities predicts $\tau_{\rm d} \propto (C_{n_e}^{2})^{2/(\beta
-2)}\ {\rm DM}^{\beta/(\beta-2)}\ \nu^{-2\beta/(\beta-2)}$, which
reduces to $\tau_{\rm d} \propto (C_{n_e}^{2})^{1.2}\, {\rm
DM}^{2.2}\, \nu^{-4.4}$ for the Kolmo\-go\-rov spectrum with $\beta=11/3$
(Romani et al.\ 1986\nocite{rnb86}).  Sutton (1971\nocite{sut71})
first noted that, for ${\rm DM}\,\la\, 20$, $\tau_{\rm d}$ increases
roughly as ${\rm DM}^{2}$, but for $20\,\la\, {\rm DM}\,\la\, 400$ the
relation steepens approaching ${\rm DM}^{4}$. With many more available
measurements of $\tau_{\rm d}$ (and $\Delta \nu_{\rm d}$) this
relation has been revisited (e.g. Rickett 1977\nocite{ric77};
Ramachandran et al.  1997\nocite{rmdm+97}). However, the basic feature
of flatter and steeper slopes for low and high DM pulsars seems to
hold well. In most of these studies the measured $\tau_{\rm d}$ or
$\Delta \nu_{\rm d}$ are scaled to a reference frequency assuming a a
priori frequency dependence of $\nu^{4.4}$. In Fig.~\ref{fig:taudm} we
show a plot of $\tau_{\rm d}$ versus DM at 400 MHz for all the pulsars
where the frequency scaling is now measured, i.e.\ all pulsars from
Fig.~\ref{fig:sidm}. We have fitted the empirical function $\tau_{\rm
d}\,({\rm ms}) = A\,{\rm DM}^{\gamma}\,(1 + B\,{\rm DM}^{\zeta})$, as
was suggested by Ramachandran et al.  (1997\nocite{rmdm+97}). Here the
term $(1 + B \,{\rm DM}^{\zeta})$ should provide a useful description
for the apparent mean dependence of the level of turbulence on DM.  In
order to assess the deviation from the Kolmogorov value we fixed
$\gamma=2.2$, and derived a best fit of the form
\begin{equation}
\tau_{\rm d}\,{\rm (ms)} = 2.5 \times 10^{-6}\,{\rm DM}^{2.2}\,(1 + 1.34 
\times 10^{-4}\,{DM}^{2.3}) 
\label{eqn:taudm}
\end{equation}
which is shown by solid line in Fig.~\ref{fig:taudm}. Comparing our
results with the fit obtained by Ramachandran et al. (1997), their
best-fit curve is slightly shifted to higher $\tau_{\rm d}$ values,
most probably caused by the large scatter of data points. The exponent
$\zeta$ is somewhat smaller than their value of 2.5, leading to a
flattening of the curve at the highest DMs. As the high DM pulsars
have frequency scaling laws which are significantly smaller than 4.4
(see Paper I), the extrapolation of their spectrum down to 400~MHz
leads to lower $\tau_{\rm d}$ values than for the case of a Kolmogorov
spectrum. This explains the flattening of the $\tau_{\rm d}$--DM curve
at the high DM end. However, the large scatter in the $\tau_{\rm d}$
values and small number statistics do not allow a tighter constraint
of $\zeta$. The different slopes for low and high DM pulsars have been
studied in detail by several investigators (e.g.\ Rickett 1977; Cordes
et al.\ 1985; Rickett 1990). An extra bias could be introduced by a
wrong conversion of $\Delta \nu_{\rm d}$ to $\tau_{\rm d}$ for the low
DM pulsars. Indeed, if $C_1\sim 1$ (see Eq.~\ref{uncertain}) does not
hold, a comparison of the $\Delta \nu_{\rm d}$ and $\tau_{\rm d}$
measurements is not possible. However, if $C_1$ differs from unity by
a constant amount, all points are shifted in the $\tau_{\rm d}$ versus
DM plot to one direction by a fixed factor. Only a systematic change
in $C_1$ with DM could change the slope of the curve itself. However,
as was discussed by Rickett (1977), theoretically there is very little
flexibility in changing $C_1$ from values close to unity. The only
experimental confirmation for this became available for the Vela
pulsar by Backer (1974\nocite{bac74}), who found $C_1=1.07$ by
extrapolating several $\tau_{\rm d}$ and $\Delta\nu_{\rm d}$
measurements at a given frequency. For PSR B1933+16, however, as
discussed earlier, it appears that there is some evidence for $C_1$
values greater than unity.

\begin{figure}
\centering
\psfig{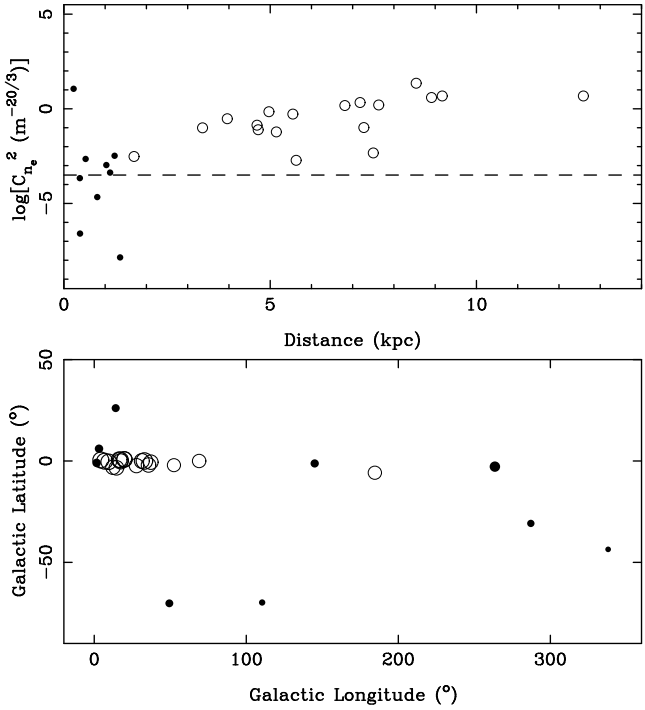}
\caption[]{ \label{fig:cne2} The top panel shows scattering strength
$\log(C_{n_e}^2)$ versus distance for 27 pulsars as in
Figure~\ref{fig:sidm}.  The dashed line corresponds to $-3.5$, the
canonical value for a homogeneous Kolmogorov medium. The bottom panel
shows the LOSs for these pulsars in the Galaxy, where the size of the
circles corresponds to the magnitude of $C_{n_e}^2$ . The filled
circles indicate decorrelation bandwidth measurements and the open
circles are pulse broadening measurements. See text for further
details.  }
\end{figure}

Alternatively, if we accept $C_1\sim1$, one can in principle explain
the steeping in the $\tau_{\rm d}$ versus DM relation by proposing a
breakdown of the homogeneity condition in the medium (Rickett 1970). A
usual way to check this condition is to analyse the scattering
strength $C_{n_e}^2$ along the LOS. For a homogeneous medium with a
Kolmogorov spectrum, $C_{n_e}^2$ can be found by using the definition
from Cordes (1986\nocite{cor86}), as
\begin{equation}
C_{n_e}^2 ({\rm m}^{-20/3}) = 0.002\,\nu^{11/3}\,D^{-11/6}\,
   \Delta\nu_{\rm d}^{-5/6}\ \ ,
\label{eqcne2}
\end{equation}
with $\nu$ in GHz, $D$ in kpc and $\Delta\nu_{\rm d}$ in MHz. We used
Eq.~\ref{eqcne2} to calculate $C_{n_e}^2$ for the pulsars from
this paper and Paper I, by averaging over all frequencies.  Our
$\tau_{\rm d}$ values were converted to $\Delta \nu_{\rm d}$ using
$C_1=1.16$, the value appropriate for a uniform medium and having a
Kolmogorov spectrum (Cordes \& Lazio 2002\nocite{cl02a}). The
distances were obtained from the Cordes \& Lazio (2002) model of the
Galactic electron density distribution, hereafter NE2001.  $C_{n_e}^2$
for other pulsars were used from Johnston et al. (1998) and Cordes et
al. (1985), but scaled appropriately for the new NE2001 model
distances.  In Fig.~\ref{fig:cne2} we plot the scattering strength
$C_{n_e}^2$ as a function of distance for all the 27 pulsars with
known frequency scaling law.  As clearly seen in the top panel,
$C_{n_e}^2$ increases with distance, a result which is consistent with
that of Cordes et al. (1985). For a homogeneous medium one expects to
find $\log(C_{n_e}^2)$ near its canonical value of $-3.5$ (e.g.\
Johnston et al.\ 1998), which is only true for the nearby pulsars.
Instead, for most of the sources with $D\,\ga\, 3$~kpc, we find
values much larger than $-3.5$ indicating that we sample highly dense
regions with enhanced scattering along these LOSs. As seen in the
bottom panel of Fig.~\ref{fig:cne2}, the LOSs with enhanced
scattering strength are concentrated in the inner regions of the
Galaxy where the majority of HII regions can be found.  These LOSs
additionally show flatter spectra of pulse broadening indicating
multiple scattering screens of anisotropic irregularities (Paper I).

%

\section{Summary \& Conclusions \label{conclusions}}

We have presented measurements of pulse broadening times for a sample
of intermediate DM pulsars. The obtained results bridge the gap in the
available literature from low DM to high DM pulsars. For both the high
DM pulsars presented in Paper I and the current sample, we find that
the PBF for a gaussian distribution of irregularities and applying the
thin screen approximation is most appropriate to describe the observed
scattering. While there are significant deviations from the expected
frequency scaling for high DM pulsars (Paper I), our current sample
is, apart from the exception for PSR B1933+16, consistent with a
Kolmogorov spectrum. Therefore, we conclude that LOSs to pulsars with
$0\,\la\, {\rm DM}\,\la\, 300$~pc cm$^{-3}$ show pulse broadening that
is consistent with a Kolmogorov spectrum of electron density
irregularities (except for a few pulsars with complicated LOSs, e.g.\
the Crab and Vela pulsars, PSR B1933+16). At around DM$=300$~pc
cm$^{-3}$ a change in the spectral index of pulse broadening is
observed leading to a flattening of the spectra. We think that this
change is related to a change of the Galactic material in the inner
region of the Galaxy. Future multi-frequency observations are highly
desirable to probe the DM range of $250-400$~pc cm$^{-3}$ and to
determine the exact transition point.

Standard theory appears still to be challenged by our result for the
relationship connecting the pulse broadening time with the
decorrelation bandwidth.  Mostly only one of these quantities is
measurable, while the other is computed using an assumed standard
relation. This can obviously lead to systematic errors which need to
be considered when combining corresponding data sets. Nevertheless,
such procedure may be still unavoidable in order to increase the size
of the studied sample, since the results obtained here and in Paper I
underline the impression that the overall state of the ISM can only be
determined in a statistical sense. Clearly, individual results are
affected by the properties of certain LOSs. For this reason, special
care is needed when interpreting the results obtained only for a small
sample of pulsars.

\begin{acknowledgements}
We thank the staff of the GMRT for help with the observations. The
GMRT is run by the National Centre for Radio Astrophysics of the Tata
Institute of Fun\-da\-men\-tal Research. We would also like to thank the
referee for useful suggestions that helped to improve the text.
\end{acknowledgements}

\bibliographystyle{aa}

\end{document}